\documentclass[aps,prl,final,reprint,showkeys,showpacs,superscriptaddress,floatfix,footinbib,twocolumn,amsmath,amssymb]{revtex4-1}

\usepackage{hyperref} 
\usepackage{url}
\usepackage{graphicx}

\usepackage{graphics}
\graphicspath{{./}}

\begin{document}

\title{Search for Cosmic-Ray Electron and Positron Anisotropies with Seven Years of Fermi Large Area Telescope Data}

\author{S.~Abdollahi}
\affiliation{Department of Physical Sciences, Hiroshima University, Higashi-Hiroshima, Hiroshima 739-8526, Japan}
\author{M.~Ackermann}
\affiliation{Deutsches Elektronen Synchrotron DESY, D-15738 Zeuthen, Germany}
\author{M.~Ajello}
\affiliation{Department of Physics and Astronomy, Clemson University, Kinard Lab of Physics, Clemson, South Carolina 29634-0978, USA}
\author{A.~Albert}
\affiliation{Los Alamos National Laboratory, Los Alamos, New Mexico 87545, USA}
\author{W.~B.~Atwood}
\affiliation{Santa Cruz Institute for Particle Physics, Department of Physics and Department of Astronomy and Astrophysics, University of California at Santa Cruz, Santa Cruz, California 95064, USA}
\author{L.~Baldini}
\affiliation{Dipartimento di Fisica ``Enrico Fermi'' dell'Universit\`a di Pisa,  I-56126 Pisa, Italy}
\affiliation{Istituto Nazionale di Fisica Nucleare, Sezione di Pisa, I-56127 Pisa, Italy}
\author{G.~Barbiellini}
\affiliation{Istituto Nazionale di Fisica Nucleare, Sezione di Trieste, I-34127 Trieste, Italy}
\affiliation{Dipartimento di Fisica, Universit\`a di Trieste, I-34127 Trieste, Italy}
\author{R.~Bellazzini}
\affiliation{Istituto Nazionale di Fisica Nucleare, Sezione di Pisa, I-56127 Pisa, Italy}
\author{E.~Bissaldi}
\affiliation{Dipartimento di Fisica ``M. Merlin" dell'Universit\`a e del Politecnico di Bari, I-70126 Bari, Italy}
\affiliation{Istituto Nazionale di Fisica Nucleare, Sezione di Bari, I-70126 Bari, Italy}
\author{E.~D.~Bloom}
\affiliation{W. W. Hansen Experimental Physics Laboratory, Kavli Institute for Particle Astrophysics and Cosmology, Department of Physics and SLAC National Accelerator Laboratory, Stanford University, Stanford, California 94305, USA}
\author{R.~Bonino}
\affiliation{Istituto Nazionale di Fisica Nucleare, Sezione di Torino, I-10125 Torino, Italy}
\affiliation{Dipartimento di Fisica, Universit\`a degli Studi di Torino, I-10125 Torino, Italy}
\author{E.~Bottacini}
\affiliation{W. W. Hansen Experimental Physics Laboratory, Kavli Institute for Particle Astrophysics and Cosmology, Department of Physics and SLAC National Accelerator Laboratory, Stanford University, Stanford, California 94305, USA}
\author{T.~J.~Brandt}
\affiliation{NASA Goddard Space Flight Center, Greenbelt, Maryland 20771, USA}
\author{P.~Bruel}
\affiliation{Laboratoire Leprince-Ringuet, \'Ecole polytechnique, CNRS/IN2P3, F-91128 Palaiseau, France}
\author{S.~Buson}
\affiliation{NASA Goddard Space Flight Center, Greenbelt, Maryland 20771, USA}
%\affiliation{NASA Postdoctoral Program Fellow, USA}
\author{M.~Caragiulo}
\affiliation{Dipartimento di Fisica ``M. Merlin" dell'Universit\`a e del Politecnico di Bari, I-70126 Bari, Italy}
\affiliation{Istituto Nazionale di Fisica Nucleare, Sezione di Bari, I-70126 Bari, Italy}
\author{E.~Cavazzuti}
\affiliation{Agenzia Spaziale Italiana (ASI) Science Data Center, I-00133 Roma, Italy}
\author{A.~Chekhtman}
\affiliation{College of Science, George Mason University, Fairfax, Virginia 22030, USA resident at Naval Research Laboratory, Washington, DC 20375, USA}
\author{S.~Ciprini}
\affiliation{Agenzia Spaziale Italiana (ASI) Science Data Center, I-00133 Roma, Italy}
\affiliation{Istituto Nazionale di Fisica Nucleare, Sezione di Perugia, I-06123 Perugia, Italy}
\author{F.~Costanza}
\email{francesco.costanza@cern.ch}
\affiliation{Istituto Nazionale di Fisica Nucleare, Sezione di Bari, I-70126 Bari, Italy}
\author{A.~Cuoco}
\affiliation{Istituto Nazionale di Fisica Nucleare, Sezione di Torino, I-10125 Torino, Italy}
\affiliation{Institute for Theoretical Particle Physics and Cosmology, (TTK), RWTH Aachen University, D-52056 Aachen, Germany}
\author{S.~Cutini}
\affiliation{Agenzia Spaziale Italiana (ASI) Science Data Center, I-00133 Roma, Italy}
\affiliation{Istituto Nazionale di Fisica Nucleare, Sezione di Perugia, I-06123 Perugia, Italy}
\author{F.~D'Ammando}
\affiliation{INAF Istituto di Radioastronomia, I-40129 Bologna, Italy}
\affiliation{Dipartimento di Astronomia, Universit\`a di Bologna, I-40127 Bologna, Italy}
\author{F.~de~Palma}
\affiliation{Istituto Nazionale di Fisica Nucleare, Sezione di Bari, I-70126 Bari, Italy}
\affiliation{Universit\`a Telematica Pegaso, Piazza Trieste e Trento, 48, I-80132 Napoli, Italy}
\author{R.~Desiante}
\affiliation{Istituto Nazionale di Fisica Nucleare, Sezione di Torino, I-10125 Torino, Italy}
\affiliation{Universit\`a di Udine, I-33100 Udine, Italy}
\author{S.~W.~Digel}
\affiliation{W. W. Hansen Experimental Physics Laboratory, Kavli Institute for Particle Astrophysics and Cosmology, Department of Physics and SLAC National Accelerator Laboratory, Stanford University, Stanford, California 94305, USA}
\author{N.~Di~Lalla}
\affiliation{Dipartimento di Fisica ``Enrico Fermi'' dell'Universit\`a di Pisa,  I-56126 Pisa, Italy}
\affiliation{Istituto Nazionale di Fisica Nucleare, Sezione di Pisa, I-56127 Pisa, Italy}
\author{M.~Di~Mauro}
\affiliation{W. W. Hansen Experimental Physics Laboratory, Kavli Institute for Particle Astrophysics and Cosmology, Department of Physics and SLAC National Accelerator Laboratory, Stanford University, Stanford, California 94305, USA}
\author{L.~Di~Venere}
\affiliation{Dipartimento di Fisica ``M. Merlin" dell'Universit\`a e del Politecnico di Bari, I-70126 Bari, Italy}
\affiliation{Istituto Nazionale di Fisica Nucleare, Sezione di Bari, I-70126 Bari, Italy}
\author{B.~Donaggio}
\affiliation{Istituto Nazionale di Fisica Nucleare, Sezione di Padova, I-35131 Padova, Italy}
\author{P.~S.~Drell}
\affiliation{W. W. Hansen Experimental Physics Laboratory, Kavli Institute for Particle Astrophysics and Cosmology, Department of Physics and SLAC National Accelerator Laboratory, Stanford University, Stanford, California 94305, USA}
\author{C.~Favuzzi}
\affiliation{Dipartimento di Fisica ``M. Merlin" dell'Universit\`a e del Politecnico di Bari, I-70126 Bari, Italy}
\affiliation{Istituto Nazionale di Fisica Nucleare, Sezione di Bari, I-70126 Bari, Italy}
\author{W.~B.~Focke}
\affiliation{W. W. Hansen Experimental Physics Laboratory, Kavli Institute for Particle Astrophysics and Cosmology, Department of Physics and SLAC National Accelerator Laboratory, Stanford University, Stanford, California 94305, USA}
\author{Y.~Fukazawa}
\affiliation{Department of Physical Sciences, Hiroshima University, Higashi-Hiroshima, Hiroshima 739-8526, Japan}
\author{S.~Funk}
\affiliation{Erlangen Centre for Astroparticle Physics, D-91058 Erlangen, Germany}
\author{P.~Fusco}
\affiliation{Dipartimento di Fisica ``M. Merlin" dell'Universit\`a e del Politecnico di Bari, I-70126 Bari, Italy}
\affiliation{Istituto Nazionale di Fisica Nucleare, Sezione di Bari, I-70126 Bari, Italy}
\author{F.~Gargano}
\affiliation{Istituto Nazionale di Fisica Nucleare, Sezione di Bari, I-70126 Bari, Italy}
\author{D.~Gasparrini}
\affiliation{Agenzia Spaziale Italiana (ASI) Science Data Center, I-00133 Roma, Italy}
\affiliation{Istituto Nazionale di Fisica Nucleare, Sezione di Perugia, I-06123 Perugia, Italy}
\author{N.~Giglietto}
\affiliation{Dipartimento di Fisica ``M. Merlin" dell'Universit\`a e del Politecnico di Bari, I-70126 Bari, Italy}
\affiliation{Istituto Nazionale di Fisica Nucleare, Sezione di Bari, I-70126 Bari, Italy}
\author{F.~Giordano}
\affiliation{Dipartimento di Fisica ``M. Merlin" dell'Universit\`a e del Politecnico di Bari, I-70126 Bari, Italy}
\affiliation{Istituto Nazionale di Fisica Nucleare, Sezione di Bari, I-70126 Bari, Italy}
\author{M.~Giroletti}
\affiliation{INAF Istituto di Radioastronomia, I-40129 Bologna, Italy}
\author{D.~Green}
\affiliation{Department of Physics and Department of Astronomy, University of Maryland, College Park, Maryland 20742, USA}
\affiliation{NASA Goddard Space Flight Center, Greenbelt, Maryland 20771, USA}
\author{S.~Guiriec}
\affiliation{NASA Goddard Space Flight Center, Greenbelt, Maryland 20771, USA}
%\affiliation{NASA Postdoctoral Program Fellow, USA}
\author{A.~K.~Harding}
\affiliation{NASA Goddard Space Flight Center, Greenbelt, Maryland 20771, USA}
\author{T.~Jogler}
\affiliation{Friedrich-Alexander-Universit\"at, Erlangen-N\"urnberg, Schlossplatz 4, 91054 Erlangen, Germany}
\author{G.~J\'ohannesson}
\affiliation{Science Institute, University of Iceland, IS-107 Reykjavik, Iceland}
\author{T.~Kamae}
\affiliation{Department of Physics, Graduate School of Science, University of Tokyo, 7-3-1 Hongo, Bunkyo-ku, Tokyo 113-0033, Japan}
\author{M.~Kuss}
\affiliation{Istituto Nazionale di Fisica Nucleare, Sezione di Pisa, I-56127 Pisa, Italy}
\author{S.~Larsson}
\affiliation{Department of Physics, KTH Royal Institute of Technology, AlbaNova, SE-106 91 Stockholm, Sweden}
\affiliation{The Oskar Klein Centre for Cosmoparticle Physics, AlbaNova, SE-106 91 Stockholm, Sweden}
\author{L.~Latronico}
\affiliation{Istituto Nazionale di Fisica Nucleare, Sezione di Torino, I-10125 Torino, Italy}
\author{J.~Li}
\affiliation{Institute of Space Sciences (IEEC-CSIC), Campus UAB, E-08193 Barcelona, Spain}
\author{F.~Longo}
\affiliation{Istituto Nazionale di Fisica Nucleare, Sezione di Trieste, I-34127 Trieste, Italy}
\affiliation{Dipartimento di Fisica, Universit\`a di Trieste, I-34127 Trieste, Italy}
\author{F.~Loparco}
\affiliation{Dipartimento di Fisica ``M. Merlin" dell'Universit\`a e del Politecnico di Bari, I-70126 Bari, Italy}
\affiliation{Istituto Nazionale di Fisica Nucleare, Sezione di Bari, I-70126 Bari, Italy}
\author{P.~Lubrano}
\affiliation{Istituto Nazionale di Fisica Nucleare, Sezione di Perugia, I-06123 Perugia, Italy}
\author{J.~D.~Magill}
\affiliation{Department of Physics and Department of Astronomy, University of Maryland, College Park, Maryland 20742, USA}
\author{D.~Malyshev}
\affiliation{Erlangen Centre for Astroparticle Physics, D-91058 Erlangen, Germany}
\author{A.~Manfreda}
\affiliation{Dipartimento di Fisica ``Enrico Fermi'' dell'Universit\`a di Pisa,  I-56126 Pisa, Italy}
\affiliation{Istituto Nazionale di Fisica Nucleare, Sezione di Pisa, I-56127 Pisa, Italy}
\author{M.~N.~Mazziotta}
\email{mazziotta@ba.infn.it}
\affiliation{Istituto Nazionale di Fisica Nucleare, Sezione di Bari, I-70126 Bari, Italy}
\author{M.~Meehan}
\affiliation{Department of Physics, University of Wisconsin-Madison, Madison, Wisconsin 53706, USA}
\author{P.~F.~Michelson}
\affiliation{W. W. Hansen Experimental Physics Laboratory, Kavli Institute for Particle Astrophysics and Cosmology, Department of Physics and SLAC National Accelerator Laboratory, Stanford University, Stanford, California 94305, USA}
\author{W.~Mitthumsiri}
\affiliation{Department of Physics, Faculty of Science, Mahidol University, Bangkok 10400, Thailand}
\author{T.~Mizuno}
\affiliation{Hiroshima Astrophysical Science Center, Hiroshima University, Higashi-Hiroshima, Hiroshima 739-8526, Japan}
\author{A.~A.~Moiseev}
\affiliation{Department of Physics and Department of Astronomy, University of Maryland, College Park, Maryland 20742, USA}
\affiliation{Center for Research and Exploration in Space Science and Technology (CRESST) and NASA Goddard Space Flight Center, Greenbelt, Maryland 20771, USA}
\author{M.~E.~Monzani}
\affiliation{W. W. Hansen Experimental Physics Laboratory, Kavli Institute for Particle Astrophysics and Cosmology, Department of Physics and SLAC National Accelerator Laboratory, Stanford University, Stanford, California 94305, USA}
\author{A.~Morselli}
\affiliation{Istituto Nazionale di Fisica Nucleare, Sezione di Roma ``Tor Vergata", I-00133 Roma, Italy}
\author{M.~Negro}
\affiliation{Istituto Nazionale di Fisica Nucleare, Sezione di Torino, I-10125 Torino, Italy}
\affiliation{Dipartimento di Fisica, Universit\`a degli Studi di Torino, I-10125 Torino, Italy}
\author{E.~Nuss}
\affiliation{Laboratoire Univers et Particules de Montpellier, Universit\'e Montpellier, CNRS/IN2P3, F-34095 Montpellier, France}
\author{T.~Ohsugi}
\affiliation{Hiroshima Astrophysical Science Center, Hiroshima University, Higashi-Hiroshima, Hiroshima 739-8526, Japan}
\author{N.~Omodei}
\affiliation{W. W. Hansen Experimental Physics Laboratory, Kavli Institute for Particle Astrophysics and Cosmology, Department of Physics and SLAC National Accelerator Laboratory, Stanford University, Stanford, California 94305, USA}
\author{D.~Paneque}
\affiliation{Max-Planck-Institut f\"ur Physik, D-80805 M\"unchen, Germany}
\author{J.~S.~Perkins}
\affiliation{NASA Goddard Space Flight Center, Greenbelt, Maryland 20771, USA}
\author{M.~Pesce-Rollins}
\affiliation{Istituto Nazionale di Fisica Nucleare, Sezione di Pisa, I-56127 Pisa, Italy}
\author{F.~Piron}
\affiliation{Laboratoire Univers et Particules de Montpellier, Universit\'e Montpellier, CNRS/IN2P3, F-34095 Montpellier, France}
\author{G.~Pivato}
\affiliation{Istituto Nazionale di Fisica Nucleare, Sezione di Pisa, I-56127 Pisa, Italy}
\author{G.~Principe}
\affiliation{Erlangen Centre for Astroparticle Physics, D-91058 Erlangen, Germany}
\author{S.~Rain\`o}
\affiliation{Dipartimento di Fisica ``M. Merlin" dell'Universit\`a e del Politecnico di Bari, I-70126 Bari, Italy}
\affiliation{Istituto Nazionale di Fisica Nucleare, Sezione di Bari, I-70126 Bari, Italy}
\author{R.~Rando}
\affiliation{Istituto Nazionale di Fisica Nucleare, Sezione di Padova, I-35131 Padova, Italy}
\affiliation{Dipartimento di Fisica e Astronomia ``G. Galilei'', Universit\`a di Padova, I-35131 Padova, Italy}
\author{M.~Razzano}
\affiliation{Istituto Nazionale di Fisica Nucleare, Sezione di Pisa, I-56127 Pisa, Italy}
%\affiliation{Funded by contract FIRB-2012-RBFR12PM1F from the Italian Ministry of Education, University and Research (MIUR)}
\author{A.~Reimer}
\affiliation{Institut f\"ur Astro- und Teilchenphysik and Institut f\"ur Theoretische Physik, Leopold-Franzens-Universit\"at Innsbruck, A-6020 Innsbruck, Austria}
\affiliation{W. W. Hansen Experimental Physics Laboratory, Kavli Institute for Particle Astrophysics and Cosmology, Department of Physics and SLAC National Accelerator Laboratory, Stanford University, Stanford, California 94305, USA}
\author{O.~Reimer}
\affiliation{Institut f\"ur Astro- und Teilchenphysik and Institut f\"ur Theoretische Physik, Leopold-Franzens-Universit\"at Innsbruck, A-6020 Innsbruck, Austria}
\affiliation{W. W. Hansen Experimental Physics Laboratory, Kavli Institute for Particle Astrophysics and Cosmology, Department of Physics and SLAC National Accelerator Laboratory, Stanford University, Stanford, California 94305, USA}
\author{C.~Sgr\`o}
\affiliation{Istituto Nazionale di Fisica Nucleare, Sezione di Pisa, I-56127 Pisa, Italy}
\author{D.~Simone}
\affiliation{Istituto Nazionale di Fisica Nucleare, Sezione di Bari, I-70126 Bari, Italy}
\author{E.~J.~Siskind}
\affiliation{NYCB Real-Time Computing Inc., Lattingtown, New York 11560-1025, USA}
\author{F.~Spada}
\affiliation{Istituto Nazionale di Fisica Nucleare, Sezione di Pisa, I-56127 Pisa, Italy}
\author{G.~Spandre}
\affiliation{Istituto Nazionale di Fisica Nucleare, Sezione di Pisa, I-56127 Pisa, Italy}
\author{P.~Spinelli}
\affiliation{Dipartimento di Fisica ``M. Merlin" dell'Universit\`a e del Politecnico di Bari, I-70126 Bari, Italy}
\affiliation{Istituto Nazionale di Fisica Nucleare, Sezione di Bari, I-70126 Bari, Italy}
\author{A.~W.~Strong}
\affiliation{Max-Planck Institut f\"ur extraterrestrische Physik, D-85748 Garching, Germany}
\author{H.~Tajima}
\affiliation{Solar-Terrestrial Environment Laboratory, Nagoya University, Nagoya 464-8601, Japan}
\affiliation{W. W. Hansen Experimental Physics Laboratory, Kavli Institute for Particle Astrophysics and Cosmology, Department of Physics and SLAC National Accelerator Laboratory, Stanford University, Stanford, California 94305, USA}
\author{J.~B.~Thayer}
\affiliation{W. W. Hansen Experimental Physics Laboratory, Kavli Institute for Particle Astrophysics and Cosmology, Department of Physics and SLAC National Accelerator Laboratory, Stanford University, Stanford, California 94305, USA}
\author{D.~F.~Torres}
\affiliation{Institute of Space Sciences (IEEC-CSIC), Campus UAB, E-08193 Barcelona, Spain}
\affiliation{Instituci\'o Catalana de Recerca i Estudis Avan\c{c}ats (ICREA), E-08010 Barcelona, Spain}
\author{E.~Troja}
\affiliation{NASA Goddard Space Flight Center, Greenbelt, Maryland 20771, USA}
\affiliation{Department of Physics and Department of Astronomy, University of Maryland, College Park, Maryland 20742, USA}
\author{J.~Vandenbroucke}
\affiliation{Department of Physics, University of Wisconsin-Madison, Madison, Wisconsin 53706, USA}
\author{G.~Zaharijas}
\affiliation{Istituto Nazionale di Fisica Nucleare, Sezione di Trieste, and Universit\`a di Trieste, I-34127 Trieste, Italy}
\affiliation{Laboratory for Astroparticle Physics, University of Nova Gorica, Vipavska 13, SI-5000 Nova Gorica, Slovenia}
\author{S.~Zimmer}
\affiliation{D\'epartement de Physique Nucl\'eaire et Corpuscolaire (DPNC), University of Geneva, CH-1211  Gen\'eve 4, Switzerland}

\collaboration{The Fermi-LAT collaboration \homepage{https://www-glast.stanford.edu/}}
\noaffiliation

\date{\today}

\begin{abstract}
The Large Area Telescope on board the Fermi Gamma-ray Space Telescope has collected the largest ever sample of high-energy cosmic-ray electron 
and positron events since the beginning of its operation. Potential anisotropies in the arrival directions of cosmic-ray 
electrons or positrons could be a signature 
of the presence of nearby sources. We use almost seven years of data with energies above 42~GeV processed with the Pass~8 
reconstruction.
The present data sample can probe dipole anisotropies down to a level of $10^{-3}$. 
We take into account systematic effects that could mimic true anisotropies at this level.
We present a detailed study of the event selection optimization
of the cosmic-ray electrons and positrons to be used for anisotropy searches. Since no significant anisotropies have been detected 
on any angular scale, we present upper limits on the dipole anisotropy. The present constraints are among the strongest to date 
probing the presence of nearby young and middle-aged sources.
\end{abstract}

% insert suggested PACS numbers in braces on next line
% insert suggested keywords - APS authors don't need to do this

%\maketitle must follow title, authors, abstract, \pacs, and \keywords

\maketitle

\section{Introduction}
\label{intro}
High-energy (GeV--TeV) charged Cosmic Rays (CRs) impinging on the top of the Earth's atmosphere 
are believed to be produced in our galaxy, most likely in Supernova Remnants (SNRs).
During their journey to our solar system, CRs are scattered on random
and irregular components of the Galactic Magnetic Field (GMF), which almost isotropize
their direction distribution. 

CR electrons and positrons (CREs) rapidly lose energy through synchrotron radiation and inverse Compton
collisions with low-energy photons of the interstellar radiation field.
As a result, CREs observed with energies of 100~GeV (1~TeV) originated from relatively nearby
locations, less than about 1.6~kpc (0.75~kpc) away~\cite{Ackermann:2010ip}; 
therefore high-energy CREs could originate from
a collection of a few nearby sources~\cite{Shen1969,Shen1970,Shen1971}.
Evidence for a local CRE source would be of great relevance for understanding 
the nature of their production.

The Large Area Telescope (LAT) on board the Fermi Gamma-ray Space Telescope observes the entire sky every 2~orbits ($\sim$3 hours)
when the satellite is operated in the usual ``sky-survey mode''~\cite{Atwood:2009ez}, making it an ideal instrument 
to search for anisotropies on any angular scale and from any direction in the sky.

In 2010, we published the results of the first 
CRE anisotropy search in the energy range above 60 GeV using the data collected by the LAT in its first year of operation, 
with null results~\cite{Ackermann:2010ip}. 
In this work, we update our previous search using the data collected  
over almost 7 years and analyzed with a new CRE event selection (Pass 8)~\cite{latcrespectrum2016}, 
in a broader energy range from 42 GeV to 2 TeV and improving the analysis methods. 

We optimized the analysis to minimize any systematic effect that could mimic a signal, for instance effects of the geomagnetic field. 
For this purpose, we performed a detailed simulation study of the usual methods for anisotropy searches to check for any possible 
features or biases on the results.
Finally, following our validation studies, we present the results obtained analyzing the LAT data, providing a sensitivity to dipole anisotropy as low as $10^{-3}$. 

\section{Analysis methods}
\label{methods}
The starting point to search for anisotropies is the construction of a reference sky map 
that should be seen by the instrument if the CRE flux was isotropic, and represents 
the null hypothesis. A comparison of the reference map 
with the actual map should reveal the presence of any anisotropies in the data. 

We perform our studies in Galactic coordinates, and we also use the zenith-centered coordinates to 
check for any feature due to the geomagnetic field.
All maps have been built using the HEALPix pixelization scheme with $N_{side}=64$~\cite{Gorski:2004by}.

Since the expected signal is tiny, four data-driven methods are used to create the reference map. These methods mitigate potential systematic 
uncertainties arising from the calculation of the detector exposure~\cite{Ackermann:2010ip}.

A set of simulated events can be generated by randomly associating detected event times and instrument angles
(``shuffling technique'' ~\cite{Ackermann:2010ip},
hereafter \textit{Method 1}). Starting from the position and orientation of the LAT at a given event time, the sky direction is recalculated using the angles 
in the LAT frame of another event randomly chosen.

An alternative method is based on the overall rate of events detected in a long time interval (``event rate technique'', hereafter \textit{Method 2}). 
Each event is assigned a time randomly chosen from 
an exponential distribution with the given average rate, and a direction  extracted from 
the actual distribution $P(\theta, \phi)$ of off-axis and azimuth angles in the LAT.
The sky direction is then evaluated using the pointing history of the LAT.
A possible issue in this method concerns the duration of the time interval chosen to calculate the average rate, since it must ensure 
adequate all-sky exposure coverage, especially in the case of a statistically limited data sample.
In fact, the presence of any small/medium angular scale anisotropies in the data would create transient fluctuations in the instantaneous 
values of $P(\theta,\phi)$ as these anisotropies pass through the LAT's field of view (FoV).
However, these anisotropies would
have no effect on average values calculated on longer time scale, since they would be averaged out~\cite{Ackermann:2010ip,Iuppa:2013pg}.

Methods 3 and 4 combine the previous techniques, i.e., one 
can extract the event time sequence from an exponential distribution with given average rate and assign the angles ($\theta$, $\phi$) 
from random events (hereafter \textit{Method 3}), or 
one can keep the observed times and draw the angles ($\theta$, $\phi$) from the distribution $P(\theta,\phi)$ (hereafter \textit{Method 4}).

We calculate the reference map by dividing the data in subsamples of two-months 
duration~\footnote{It is worth mentioning that the precession period of the Fermi orbit is 55 days.}, 
then we add the maps corresponding to each period. Such choice guarantees 
averaging intervals that are long enough to smear out possible medium/large scale anisotropies but, at the same time, 
that are short compared to changing data-taking conditions (i.e. solar cycle, any change in the LAT performance, etc.). 

Once the reference map is known, a simple pixel-to-pixel comparison with the real map can be performed to search for 
statistically significant deviations.
This method is indeed applied to integrated sky maps, in which each pixel contains the integrated number of events in a given
circular region around the pixel itself. In case of an anisotropy with angular scale similar to the integration region, spillover effects are reduced increasing sensitivity~\cite{Ackermann:2010ip}. 

Another strategy is the spherical harmonic analysis of a fluctuation sky map.
The fluctuation in each pixel is defined as 
$f_i = n_i / \mu_i-1$, 
where $n_i$ ($\mu_i$) is the number of events in the $i-th$ pixel in the real (reference) map.
The fluctuations map is expanded in the basis of
spherical harmonics, producing a set of coefficients $a_{lm}$, used to build the auto angular
power spectrum (APS)
$\hat{C}_l=\sum^{l}_{m=-l}|a_{lm}|^2 / (2l+1).$
An increased power $\hat{C}_{l}$ at a multipole $l$ corresponds to an anisotropic excess at angular scale $\sim180^{\circ}/l$.

Any deviation of the APS from Poisson noise $C_N$ will be a hint of anisotropies. The Poisson noise (also known as white or shot noise) is due 
to the finite number of events in the map, so that $\hat{C}_l = C_N + \hat{C}_{l}^{ani}$. 
To check whether the observed power spectrum $\hat{C}_{l}$ is statistically compatible with the Poisson noise, we tested the null hypothesis 
$\hat{C}_{l}=C_{N}$ against the alternative one $\hat{C}_l = C_N + \hat{C}_{l}^{ani}$, with 
$\hat{C}_{l}^{ani}>0$. 
The white noise over a full sky observed with uniform exposure is $C_{N}=4\pi / N$,
where $N$ is the total number of observed events. To account for a non-uniform exposure map, the white noise
is given by 
$C_{N}= (4\pi/ N_{pixels}^2) \sum_{i=1}^{N_{pixels}} n_i / \mu_i^2$~\cite{Fornasa2016}.

\section{Event selection}
\label{optcut}

We select time intervals (Good Time Intervals, GTIs) when the LAT is operating in standard sky survey mode outside the South 
Atlantic Anomaly (SAA) and removing the times when the LAT is oriented at rocking angles
exceeding $52^{\circ}$~\footnote{Starting from September 2009, the Fermi telescope operated in survey mode with a rocking angle of $50^{\circ}$ 
(against $35^{\circ}$ in the first year of operation).}. 

Assuming an isotropic distribution of CREs at very large distances from the Earth, not all 
of these particles are able to reach the LAT due to the geomagnetic field and Earth's occultation. 
In the case of CREs there are regions where only positrons or 
electrons are allowed (in the West and in the East, respectively)~\cite{FermiLAT:2011ab}.
Therefore, a dedicated selection is employed by means of simulations to reduce the geomagnetic effects on the arrival directions of CREs detected by the LAT.
We summarize the results in the next section and the details of our studies are given in Supplementary  
Material~\footnote{See Supplemental Material at the link \url{http://link.aps.org/supplemental/10.1103/PhysRevLett.118.091103} 
for further information about the data analysis method, the procedure used to optimize the event 
selection and the calculation of the upper limit of the dipole anisotropy, which includes Refs.~\cite{Campbell:2014mpa,Li:1983fv}.}.

\section{Validation studies}
\label{valid}
To check the analysis methods and the prediction for the noise of the APS, we developed a simulation of an ideal detector with a 
FoV radius ranging from $40^{\circ}$ to $180^{\circ}$, which includes 
the real spacecraft position, orientation and livetime of the LAT~\footnote{Different FoVs will create different levels of non-uniformity in 
the exposure map.}. We performed 1000 independent realizations with an isotropic
event distribution at a rate of 0.1 Hz, covering the same time interval of the current analysis. 
The simulated event samples are analyzed with the same chain as the real one, and with the same GTI selections described above.

We used the 1000 simulated data sets to check the four analysis methods discussed above. 
For each realization we applied each method 25 times and we calculated the average reference map.
Then we calculated the APS 
with the {\tt anafast} code ~\cite{Gorski:2004by}
by comparing each simulated map with the corresponding reference map.

\begin{figure}[!ht]
\includegraphics[width=1\columnwidth,height=0.2\textheight,clip]{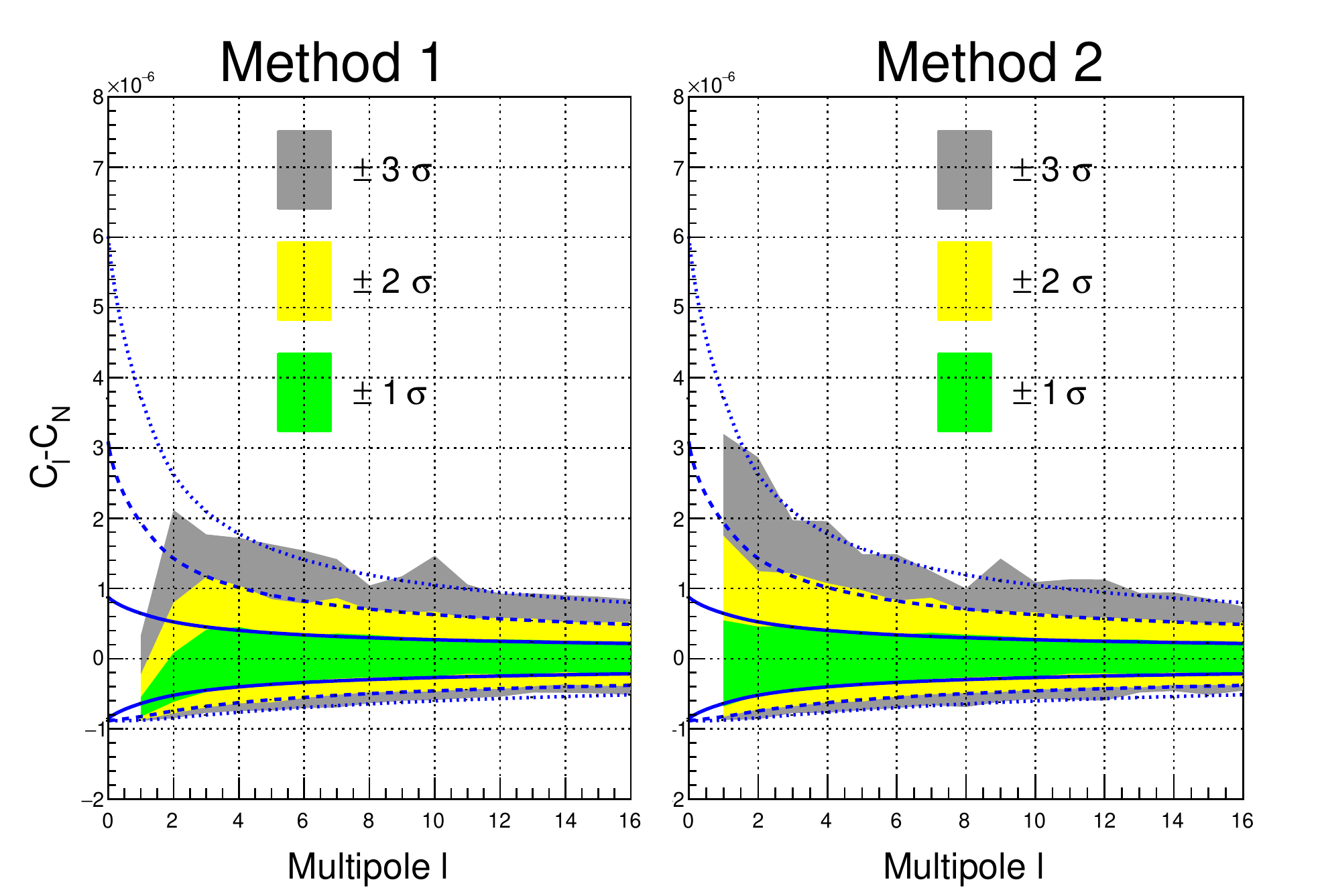}
\caption{Method 1 (left) and Method 2 (right) APS as a function of the multipole $l$ for ideal detectors with a $50^{\circ}$ FoV radius based on 1000
independent simulations.
The colored bands show the regions corresponding to different quantiles at $\pm 1 \sigma$ (green), $\pm 2 \sigma$ (yellow) and $\pm 3 \sigma$ (gray) respectively.
The blue lines show the calculation from the white noise distribution at the same quantile values. The fluctuations outside the $2 \sigma$
region are due to the limited number of simulations.}
\label{Fig2apsSimu}
\end{figure}

Figure~\ref{Fig2apsSimu} shows an example of the APS obtained using Methods 1 and 2 for the case of an ideal detector with $50^{\circ}$ FoV radius.
Further details of this study are presented in Supplementary Material,
and the results can be summarized as follows: i) all the methods give the same white noise value:
ii) Methods 1 and 4 show some bias with respect to (\mbox{w.r.t.}) the white noise level at low multipoles, comparable with the angular scale of the 
FoV; iii) Methods 2 and 3 show a better behavior \mbox{w.r.t.} the white noise value.
As discussed above, the shuffling technique is based on an event time sequences fixed to the real one, and this can 
break the Poisson random process between events on an angular scale larger than the FoV.

We performed an additional simulation injecting a dipole anisotropy from the direction $(l=230^{\circ} ,b=-3^{\circ})$ with different amplitudes ranging 
between 10\% and 0.1\% (expected sensitivity limit due to the statistics).
We were able to detect these anisotropies with the shuffling and rate methods in the case of large anisotropy 
amplitude \mbox{w.r.t.} the sensitivity limit. 
However, the true dipole anisotropy is underestimated, in particular with the shuffling method.
Further details on this validation study can be found in Supplementary Material.

Finally, we performed a further validation study based on the CRE LAT Instrument Response Functions (IRFs) for electrons
and protons (which contaminate the CRE sample). We simulated an isotropic distribution with electron, positron and proton 
intensities according to the AMS02 data~\cite{Aguilar:2014mma,Aguilar:2015ooa}, still using the real attitude of the spacecraft with the real LAT livetime.
The geomagnetic effects were also taken into account by back-tracking each primary particle from the LAT to 10 Earth radii, 
to check if it can escape (allowed direction), or if it intercepts the Earth or it is trapped in the geomagnetic field (forbidden direction).
We used the International Geomagnetic Reference Field model (IGRF-12)~\cite{Thébault2015} to describe the magnetic field in the proximity of the Earth.

We performed the analysis in nine independent energy bins from 42 GeV to 2 TeV. 
To reduce the geomagnetic effects below the level of our sensitivity, we performed the analysis with a reduced FoV, i.e., 
we set the allowed maximum off-axis
angle as a function of energy. As a result, the maximum zenith angle that could be observed is set by the FoV, since the angle between the
LAT Z-axis (on-axis direction) and the zenith (i.e., the rocking angle) is fixed with the sky-survey attitude.

\begin{figure}[!ht]
\includegraphics[width=1\columnwidth,height=0.2\textheight,clip]{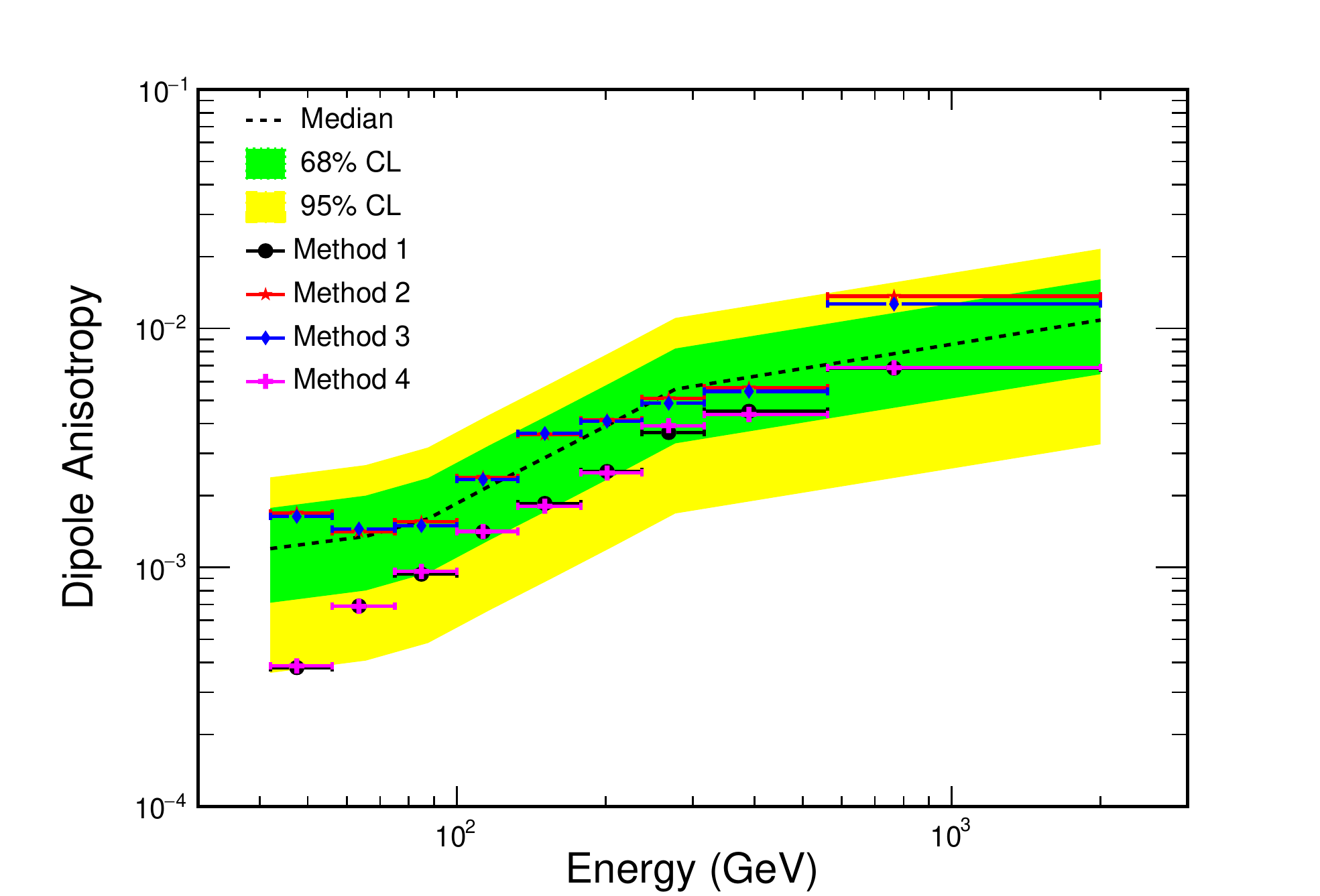}
\includegraphics[width=1\columnwidth,height=0.2\textheight,clip]{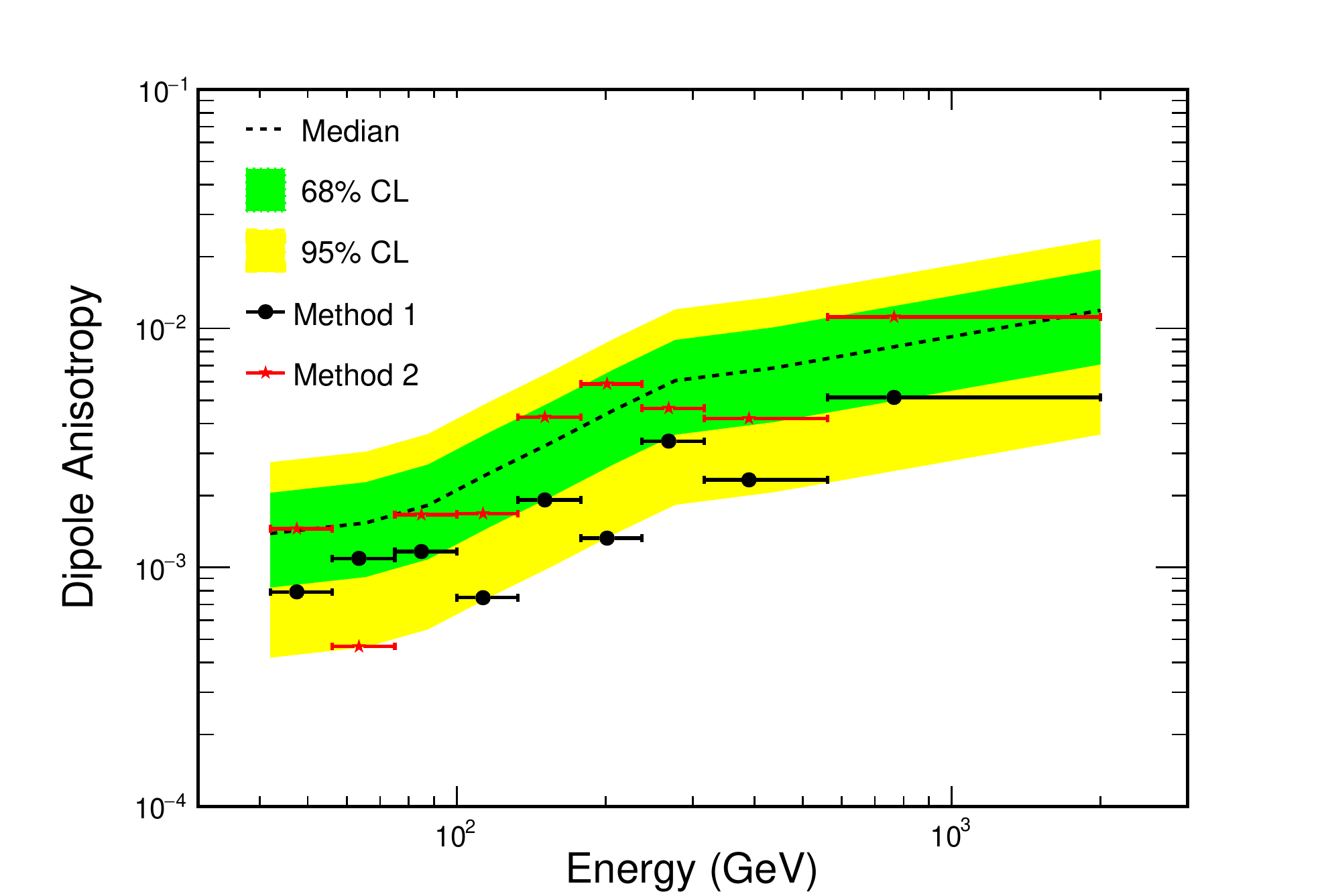}
\caption{Dipole anisotropies as a function of energy. 
Top panel: simulated isotropic data using Methods 1-4. 
Bottom panel: real data using Methods 1-2.
The markers (median energy value
calculated for a power-law flux with a spectral index of \mbox{-3}) show the results and 
the horizontal error bars indicate the energy bin width. 
The colored bands show the expected central confidence intervals of the white-noise
at 68\% and at 95\%.}
\label{Figdelta}
\end{figure}

We adopt this strategy to avoid any distortion of the distribution of arrival directions in the instrument 
coordinates, since in the analysis we assume that this distribution is the same as the one generated by an 
isotropic arrival distribution. The final set of maximum off-axis ($\theta$) angles are: 
$\theta<40^{\circ}$ for E(GeV) in the range [42, 56]; $\theta<50^{\circ}$ for E(GeV) in the range [56, 75] and $\theta<60^{\circ}$ for E(GeV)$>$75.
The maximum off-axis angle used in the current work corresponds to the one used to reconstruct the LAT CRE spectrum~\cite{latcrespectrum2016}. 

We calculate the APS with the four methods introduced above. For each method we average 10000 realizations to create 
the reference map to be used to extract the APS. 

Figure~\ref{Figdelta} shows the dipole anisotropy $\delta=3\sqrt{C_1 / 4\pi}$~\cite{Ackermann:2010ip} calculated using the $C_1$ values 
as a function of energy for the last simulation for Methods 1-4 (top panel).
The colored bands show the expected confidence intervals due to the white noise, i.e., assuming the null hypothesis $\hat{C}_{l}^{ani}=0$,
and correspond to the 68\% and 95\% central confidence intervals of $\delta$.
Methods 1 and 4 underestimate the white noise level, in particular for the low energy bins 
(i.e., those with smaller FoVs), still in the expected band, while Method 2 and 3 show a better behavior.
These results are similar to those discussed in the case of ideal detectors with different FoVs.
Further details are discussed in the SOM.
Given the compatibility of the results of Method 1 with 4 and Method 2 with 3 we decided to analyze data using only 
Method 1 and 2~\footnote{We decided to use these two methods for a consistency check despite
introducing many trials in the analysis chain and reducing the post-trials significance in case of any detections.}.

\section{Data analysis and discussion}
We performed the analysis on real data in nine independent energy bins with  
energy-dependent FoVs as discussed above, on a total of about 12.2M (52k) of events above 42 (562) GeV.

We present in the SOM
the maps for the various energy bins in zenith-centered and Galactic coordinates. 
We also show the significance maps in Galactic coordinates obtained by comparing
the integrated reference maps produced with Method 2 to the actual integrated maps. 
The significances shown in these maps are pre-trials, 
i.e., they do not take into account the correlations between adjacent pixels (see.~\cite{Ackermann:2010ip} for a full discussion).
In any case, none of these maps indicates significant excesses or deficits at any angular scale, showing that our measurements
are consistent with an isotropic sky. 

\begin{figure}[!ht]
\includegraphics[width=1\columnwidth,height=0.2\textheight,clip]{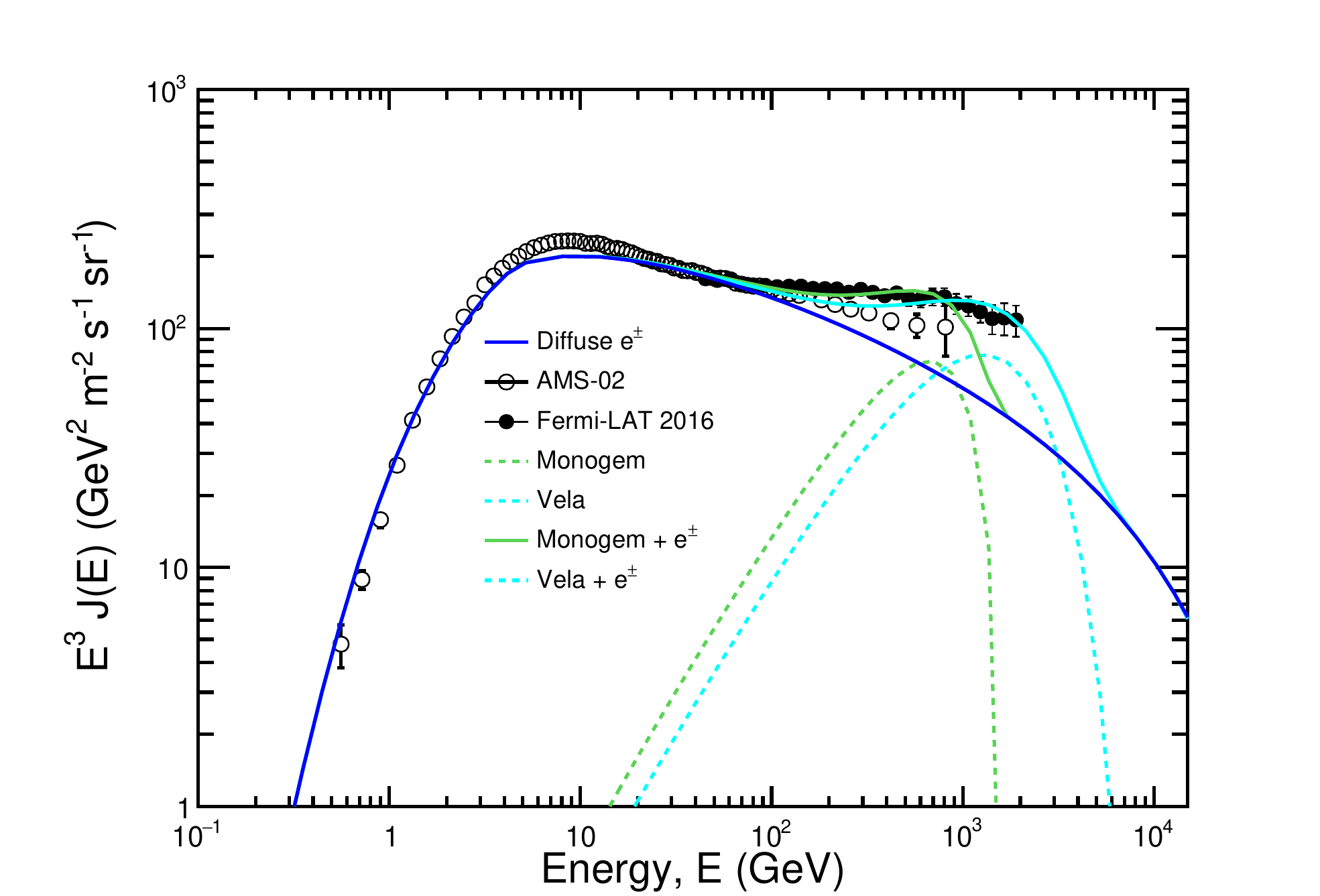}
\includegraphics[width=1\columnwidth,height=0.2\textheight,clip]{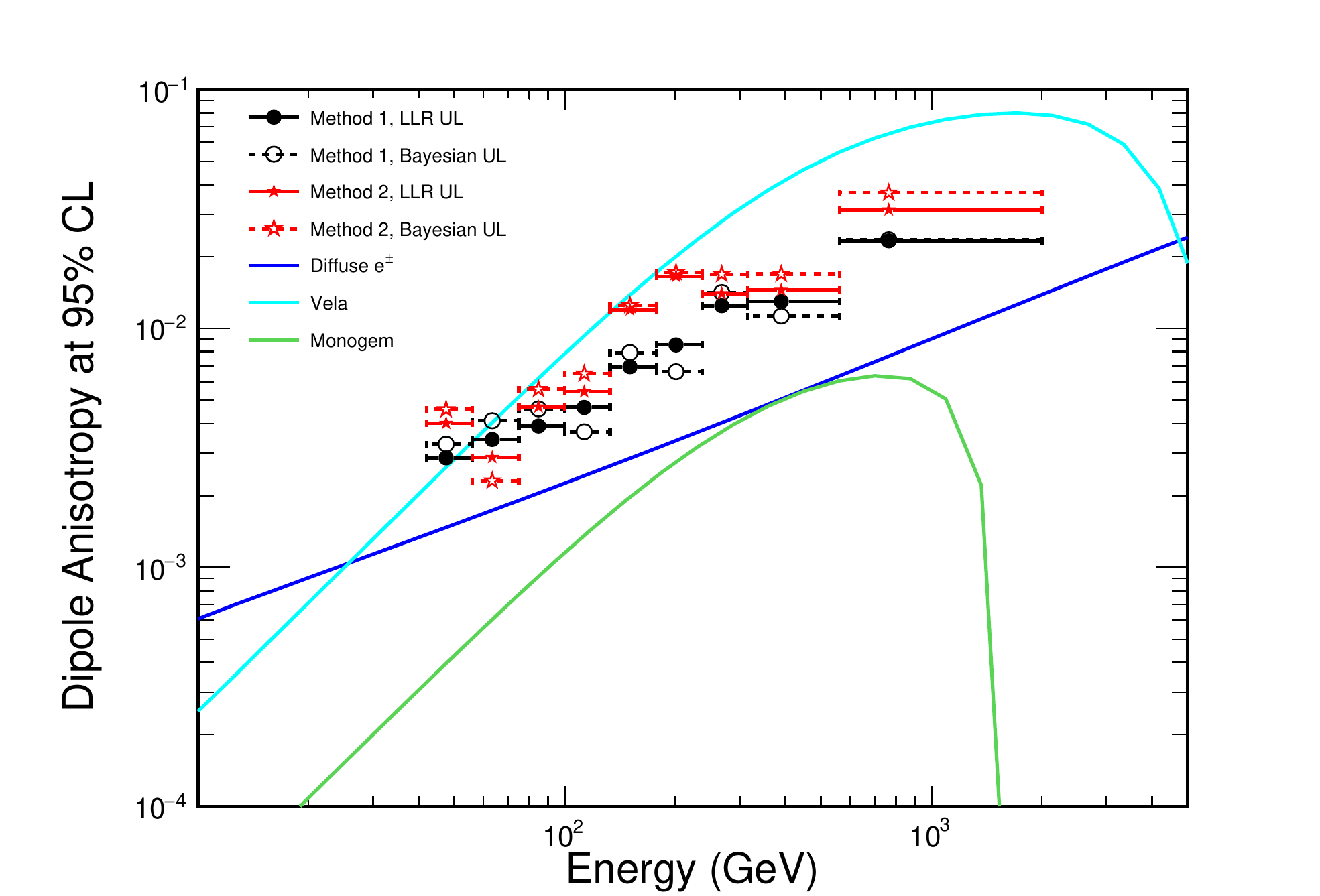}
\caption{Top panel: CRE spectra measured by the Fermi-LAT~\cite{latcrespectrum2016} and AMS02~\cite{Aguilar:2014mma}.
Blue line: Galactic CRE evaluated with {\tt DRAGON-FLUKA}~\cite{Mazziotta:2015uba};
Green line: Monogem alone (dashed) and total (solid). Cyan line: Vela alone (dashed) and total (solid).
Bottom panel: Upper limit at 95\% CL on dipole anisotropies as a function of energy. 
The markers in this panel show the actual measurements.}
\label{Figdeltaul}
\end{figure}

We have calculated the APS for real data with Method 1 and 2 
for the nine energy bins (see Figs. [15] and [16] of the SOM). The current results lie within the 
3$\sigma$ range of the expected white noise up to angular scale of a few degrees, showing the 
consistency with an isotropic sky for all energy bins tested and for $l < 30$.
In particular, Fig.~\ref{Figdelta} (bottom panel) shows the dipole anisotropy as a function of energy calculated from the $C_1$ 
evaluated with Methods 1 and 2. Since no significant anisotropies have been detected, 
we calculate upper limits on the dipole anisotropy (Fig.~\ref{Figdeltaul}).

The current results can be compared with the expected anisotropy from Galactic CREs. 
Figure~\ref{Figdeltaul} (top panel) shows the spectrum of the Galactic CREs 
component evaluated with the {\tt DRAGON} propagation code
(2D version)~\cite{Evoli:2008dv} with secondary particles production from Ref.~\cite{Mazziotta:2015uba}, assuming that
the scalar diffusion coefficient depends on the particle rigidity $R$ and on the distance from the Galactic plane $z$ according 
to the parametrization $D \,\, = \,\, D_0 \, \left( R / R_0 \right)^{0.33} \, e^{|z|/z_t}$,
where $D_0=4.25 \times 10^{28}~\textrm{cm}^2 {\textrm s}^{-1}$, $R_0 =$ 4 GV and $z_t = $ 4 kpc. 
The Alfv\'en velocity is set to $v_{\rm A} = 33~\textrm{km}~\textrm{s}^{-1}$.
In the same figure, the intensity expected from individual sources located in the Vela (290~pc distance and 1.1$\times 10^4$~yr age) 
and Monogem (290~pc distance and $1.1\times10^5$~yr age) positions are also shown. For the single sources, we have adopted a burst-like 
electron injection spectrum in which the duration of the emission is much shorter than the travel 
time from the source, described by a power law with index $\Gamma=1.7$ and with an exponential cut-off $E_{cut}$=1.1~TeV, i.e., 
$Q(E)=Q_0~E(\textrm {GeV})^{-\Gamma}~\exp(-E/E_{cut})$ (see Refs.~\cite{Ackermann:2010ip,Grasso:2009ma})~\footnote{The solar
modulation was treated using the force-field approximation with $\Phi$=0.62~GV~\cite{Gleeson:1968zza}.}.
For both sources, the value of the normalization constant $Q_0$ has been chosen to obtain a total flux not higher than 
that measured by the Fermi-LAT~\cite{latcrespectrum2016} and by AMS02~\cite{Aguilar:2014mma}. 
Possible effects of the regular magnetic field on the predicted dipole are not considered here (see~\cite{Ahlers:2016njd}).

Figure~\ref{Figdeltaul} shows the upper limits (UL) at 95\% CL on the dipole anisotropy $\delta$ as a function of energy.
We calculate the ULs using the frequentist (log-likelihood ratio, LLR) and Bayesian methods.
The current ULs as a funtion of energy at 95\%CL range from $\sim 3 \times 10^{-3}$ to $\sim 3 \times 10^{-2}$, of a factor 
of about 3 better than the previous results~\footnote{We point out that 
the current limits on the dipole 
anisotropies have been calculated in independent energy bins, while in Ref.~\cite{Ackermann:2010ip} they were calculated as a function of  
minimum energy. In addition, the current methodology to calculate the ULs is different \mbox{w.r.t.} the 
one used in Ref.~\cite{Ackermann:2010ip} (see SOM)}.
 
In Fig.~\ref{Figdeltaul} the anisotropy due to the Galactic CREs is also shown, together with the one
expected from Vela and Monogem sources based on the same models used for estimating potential spectral contributions from them~\cite{Grasso:2009ma}.
The current limits on the dipole anisotropy are probing nearby young and middle-aged sources.

The current results on the CRE anisotropy with the measurements of their spectra can constrain the production of 
these particles in Supernova Remnants and Pulsar Wind Nebulae~\cite{DiBernardo:2010is,Linden:2013mqa,Manconi:2016byt} or from dark matter 
annihilation~\cite{Cernuda:2009kk,Borriello:2010qh}.

The Heliospheric Magnetic Field (HMF) can also affect the directions of CREs, but 
it is not easy to quantify its effect. 
A dedicated analysis in ecliptic coordinates would be sensitive to HMF effects. 
However, such analysis was performed with 1 year of CRE data above 60 GeV to constrain dark matter models 
without finding any significant feature~\cite{Ajello:2011dq}.

Anisotropy that is not associated with the direction to nearby CR sources is expected
to result from the Compton-Getting (CG) effect \cite{CG1935}, in which 
the relative motion of the observer w.r.t the CR plasma changes the 
intensity of the CR fluxes, with larger intensity arriving from the direction of
motion and lower intensity arriving from the opposite direction. 
The expected amplitude of these motions is less than $10^{-3}$, smaller than the sensitivity of this search. 

Contamination of the CRE sample with other species (protons) can introduce some systematic uncertainties in the measurement.
Ground experiments have detected anisotropies for protons of energies above 10~TeV at the $10^{-3}$ level. 
These anisotropies decrease with decreasing energies, and since the proton contamination in our CRE selection is 
about 10\%~\cite{latcrespectrum2016}, the total
anisotropy from proton contamination is expected to be less than $10^{-4}$, much smaller than the current sensitivity.
Moreover, being \mbox{$\delta \sim 1/\sqrt{N}$}, including the proton contamination would increase the measured 
limits by a factor \mbox{$\sim 1/\sqrt{1-\alpha_p}$}, where $\alpha_p$ is the contamination. Such an increase would 
be noticeable only in the highest-energy bin and can be quantified to $\sim 5\%$.

% If you have acknowledgments, this puts in the proper section head. 
\begin{acknowledgments} 
\section{Acknowledgments} 
The Fermi LAT Collaboration acknowledges generous ongoing support
from a number of agencies and institutes that have supported both the
development and the operation of the LAT as well as scientific data analysis.
These include the National Aeronautics and Space Administration and the
Department of Energy in the United States, the Commissariat \`a l'Energie Atomique
and the Centre National de la Recherche Scientifique / Institut National de Physique
Nucl\'eaire et de Physique des Particules in France, the Agenzia Spaziale Italiana
and the Istituto Nazionale di Fisica Nucleare in Italy, the Ministry of Education,
Culture, Sports, Science and Technology (MEXT), High Energy Accelerator Research
Organization (KEK) and Japan Aerospace Exploration Agency (JAXA) in Japan, and
the K.~A.~Wallenberg Foundation, the Swedish Research Council and the
Swedish National Space Board in Sweden.Additional support for science analysis during the 
operations phase is gratefully acknowledged from the Istituto Nazionale di Astrofisica in Italy and the Centre 
National d'\'Etudes Spatiales in France.

The authors acknowledge the use of HEALPix \url{http://healpix.sourceforge.net} described in K.M. Gorski et al., 2005, Ap.J., 622, p.759.
S.~B. and S.~G acknowledges support as a NASA Postdoctoral Program Fellow, USA. M.R. acknowledges
funded by contract FIRB-2012-RBFR12PM1F from the Italian Ministry of Education, University and Research (MIUR).

The authors acknowledge the use of HEALPix \url{http://healpix.sourceforge.net} described in K.M. Gorski et al., 2005, Ap.J., 622, p.759
\end{acknowledgments}
% 

% Create the reference section using BibTeX:

%\bibliographystyle{unsort}

\bibliographystyle{apsrev4-1}
\bibliography{LATCREAnis2.bib}{}

%\appendix

%\input{LATCREAnis2_SOM_PRLReSub}
%\nextpage

\end{document}